\documentclass[conference]{IEEEtran}
\IEEEoverridecommandlockouts
\usepackage{balance}
\usepackage{cite}
\usepackage{float}
\usepackage{amsmath,amssymb,amsfonts}
\usepackage{graphicx}
\usepackage{subcaption}
\usepackage{textcomp}
\usepackage{afterpage}
\usepackage{adjustbox}
\usepackage{stfloats}  
\usepackage{enumitem}
\usepackage{xcolor}
\usepackage{soul}
\sethlcolor{yellow}
\usepackage{tabularx,ragged2e,booktabs}
\usepackage{algorithm}
\usepackage{hyperref}
\usepackage{algorithm,algorithmicx}
\usepackage{algpseudocode}
\usepackage{xfrac}
\usepackage{url}
\usepackage{microtype}
\usepackage{afterpage}
\usepackage{amstext} 
\usepackage{array}   
\newcolumntype{L}{>{$}l<{$}}

\newcolumntype{C}{>{\raggedright\arraybackslash}X}

\newlength\mylen
\newcolumntype{M}{>{\RaggedRight}m{\mylen}}
\usepackage[switch]{lineno}
\usepackage[T1]{fontenc}
\usepackage{tcolorbox}
\usepackage[scaled=0.85]{beramono}
\usepackage{fancyhdr}
\makeatletter
\algrenewcommand\ALG@beginalgorithmic{\footnotesize}
\algrenewcommand\algorithmiccomment[2][\footnotesize]{{#1\hfill\(\triangleright\) #2}}
\makeatother

\renewcommand{\Comment}[1]{// {\fontfamily{fvm}\selectfont #1}}

\def\BibTeX{{\rm B\kern-.05em{\sc i\kern-.025em b}\kern-.08em
    T\kern-.1667em\lower.7ex\hbox{E}\kern-.125emX}}

\fancypagestyle{alim}{\fancyhf{}\fancyfoot[L]{\footnotesize \textit{The paper has been accepted at the 21st IEEE International Conference on Software Architecture (ICSA) 2024}}}

\begin{document}

\title{Smart HPA: A Resource-Efficient Horizontal Pod Auto-scaler for Microservice Architectures \vspace{-0.5em}}

\author{%
    \begin{minipage}{1.0\textwidth}
        \centering
        Hussain Ahmad$^{\ast}$,
        Christoph Treude$^{\dagger}$,
        Markus Wagner$^{\S}$,
        Claudia Szabo$^{\ast}$ \\
        \smallskip
        $^{\ast}$University of Adelaide, Australia,
        $^{\dagger}$University of Melbourne, Australia,
        $^{\S}$Monash University, Australia \\
        \{hussain.ahmad, claudia.szabo\}@adelaide.edu.au, christoph.treude@unimelb.edu.au, markus.wagner@monash.edu
\vspace{-1.0\baselineskip}
    \end{minipage}
}

\maketitle

\begin{abstract}
Microservice architectures have gained prominence in both academia and industry, offering enhanced agility, reusability, and scalability. To simplify scaling operations in microservice architectures, container orchestration platforms such as Kubernetes feature Horizontal Pod Auto-scalers (HPAs) designed to adjust the resources of microservices to accommodate fluctuating workloads. However, existing HPAs are not suitable for resource-constrained environments, as they make scaling decisions based on the individual resource capacities of microservices, leading to service unavailability and performance degradation. Furthermore, HPA architectures exhibit several issues, including inefficient data processing and a lack of coordinated scaling operations. To address these concerns, we propose Smart HPA, a flexible resource-efficient horizontal pod auto-scaler. It features a hierarchical architecture that integrates both centralized and decentralized architectural styles to leverage their respective strengths while addressing their limitations. We introduce resource-efficient heuristics that empower Smart HPA to exchange resources among microservices, facilitating effective auto-scaling of microservices in resource-constrained environments. Our experimental results show that Smart HPA outperforms the Kubernetes baseline HPA by reducing resource overutilization, overprovisioning, and underprovisioning while increasing resource allocation to microservice applications.

\end{abstract}

\begin{IEEEkeywords}
Microservices, Auto-scaling, Self-Adaptation, Software Architecture, Resource Management, Kubernetes
\end{IEEEkeywords}

\sloppy

\section{Introduction} \label{section1}

Microservice architectures have gained widespread popularity in recent years \cite{pimentel2021self}, with several prominent enterprises (e.g., Netflix, Amazon, and Spotify) and defense systems transitioning from monolithic to microservices for improving their service quality \cite{rossi2022dynamic, ahmad2023review}. In the microservice software design paradigm, complex applications are broken down into a set of autonomous, loosely coupled, and fine-grained services called microservices. Each microservice has a defined functionality and communicates with others through lightweight interfaces such as the HTTP resource API \cite{pimentel2021self}. Microservice architectures accelerate application delivery and improve reliability, as each microservice is designed, developed, and deployed independently \cite{gorden2023predicting}. In general, microservices are deployed using software containers, with container orchestration platforms being widely adopted for the runtime management of microservices \cite{rossi2022dynamic}. While various container orchestration platforms such as Kubernetes \cite{dobies2020kubernetes}, Docker Swarm \cite{soppelsa2016native}, and Red Hat OpenShift \cite{dumpleton2018deploying}, are available \cite{zhou2022containerization}, Kubernetes is the most widely adopted platform in both academia and industry \cite{rossi2020hierarchical}.

Auto-scaling is an essential requirement for microservice architectures as it allows a microservice to dynamically adjust its computing resources to handle fluctuating workloads (e.g., Slashdot effect \cite{liu2022coordinating}) without human intervention \cite{zargarazad2023auto}. For instance, during holiday seasons such as Black Friday, websites like Amazon experience a tenfold increase in workload compared to their usual levels \cite{bodik2010characterizing}. This increased load depletes the allocated resources for active microservices, resulting in longer response times and service unavailability. To address this challenge, container orchestration platforms include a Horizontal Pod Auto-scaler (HPA) that automatically adjusts the number of replica pods within a microservice deployment in response to fluctuating workloads \cite{nguyen2020horizontal}. 

HPAs consider workload fluctuation, resource utilization, and user requirements specified in a Service-Level Agreement (SLA) \cite{abdel2023intelligent} to determine the required number of microservice replicas in dynamic environments. However, existing HPAs have a number of limitations, including the inability to scale a microservice deployment beyond the predefined resource limits \cite{nguyen2020horizontal}, and the vulnerability of their architectures to failures \cite{rossi2020hierarchical}. In the context of microservice deployment, each microservice within an application is allocated specific resources (i.e., maximum replica limit) \cite{zargarazad2023auto, nguyen2020horizontal} and HPAs are bound by these predefined maximum replica limits. Hence, when a microservice application experiences a high workload, HPAs are unable to scale busy microservices within the application beyond their maximum replica limits \cite{nguyen2020horizontal}. This limitation, in turn, leads to service unavailability, performance degradation, and financial losses \cite{baarzi2021showar}. At the same time, less busy microservices in the application have surplus resources, leading to resource wastage and additional operational costs \cite{yu2020microscaler}. In addition, HPA control architectures have been implemented either in a centralized \cite{gias2019atom} or a decentralized \cite{nitto2020autonomic} manner, resulting in a single point of failure \cite{rossi2020geo} or a lack of coordination among scaling decisions \cite{rossi2020hierarchical}, respectively. Recent improvements to decentralized solutions have been proposed through master-worker \cite{imdoukh2020machine} and hierarchical \cite{rossi2020self} architectures. However, these improvements have not effectively addressed the communication overhead challenges \cite{weyns2013patterns}.

\thispagestyle{alim}

To address these challenges, we propose Smart HPA as an extension of Kubernetes. Smart HPA incorporates a hierarchical architecture that combines centralized and decentralized architectural styles. The decentralized architecture of Smart HPA consists of dedicated auto-scalers for each microservice that handle scaling operations for their respective microservices. The centralized component of Smart HPA activates only in resource-constrained scenarios, facilitating resource exchange among microservices to enable resource-efficient scaling of microservices. Through this deliberate restriction of communication interactions between decentralized and centralized components, Smart HPA reduces communication overhead compared to traditional hybrid approaches, as the extent of communication overhead depends on these interactions \cite{weyns2013patterns}. Furthermore, through the utilization of decentralized auto-scalers, the proposed hierarchical architecture mitigates challenges associated with centralized architectures, such as the vulnerability of a single point of failure. In addition, Smart HPA provides flexibility regarding the scaling policies employed by dedicated auto-scalers and the scaling metrics utilized, such as CPU usage and response time, for executing auto-scaling of microservices. This flexibility caters to the needs of researchers and practitioners by enabling them to tailor scaling policies and metrics according to their requirements. In summary, our contributions are three-fold.

\begin{itemize}[leftmargin=*]

    \item We propose a hierarchical architecture that integrates both centralized and decentralized architectural styles to capitalize on their advantages while addressing limitations in managing auto-scaling operations for microservice applications. 
    
    \item  We develop resource-efficient heuristics that transfer resources from overprovisioned to underprovisioned microservices within a microservice application, facilitating scaling operations in resource-constrained environments.
    
    \item We evaluate the performance of Smart HPA against the widely used Kubernetes HPA using a real-world microservice benchmark application. With the default configurations of the benchmark application, our experimental results show that Smart HPA excels with 5x less overutilization, 7x lower overprovisioning, no underprovisioning, and a 1.8x boost in microservice resource allocation.
     
\end{itemize}

\noindent We provide the replication package \cite{replication} for Smart HPA, encompassing all scripts and data essential for reproducing, validating, and extending the results outlined in the paper.

\section{Background and Related Work} \label{section5}

In this section, we investigate existing scaling policies and architectural designs of HPAs.

\vspace{-0.2cm} \subsection{Scaling Policies}
\noindent\textit{i) Threshold-based scaling policy.} Threshold-based policies are widely used in both academia (e.g., \cite{hossen2022lightweight, balla2020adaptive, nitto2020autonomic, al2017autonomic}) and industrial container orchestration platforms (e.g., Kubernetes, Amazon ECS) to determine desired replica counts of microservices using different scaling metrics such as CPU and memory usage \cite{santos2023gym}. A threshold-based scaling policy is composed of conditions that trigger scaling actions, such as "If the CPU usage exceeds 80 percent, then scale up the microservice". While threshold-based policies are straightforward in concept, they require manual threshold adjustment for dynamic auto-scaling of microservices \cite{yu2020microscaler}. Furthermore, defining threshold-based policies for complex microservice infrastructures with resource contention and inter-dependencies is challenging \cite{rossi2020self}.
\newline \noindent\textit{ii) Fuzzy-based scaling policy.} A fuzzy-based scaling policy refers to a predetermined set of "If-Else" rules that rely on human knowledge of microservice applications to make scaling decisions \cite{liu2018fuzzy, persico2017fuzzy, arabnejad2017comparison}. Fuzzy inference, in contrast to threshold-based policies, permits the use of descriptive terms (e.g., high, medium, low) rather than exact numbers when specifying rules for scaling decisions \cite{qu2018auto}. Fuzzy-based policies, like threshold-based scaling, demand a comprehensive understanding of the application for defining fuzzy rules \cite{rossi2020self, yu2020microscaler}.
\newline \noindent\textit{iii) Queuing theory-based scaling policy.} The existing literature (e.g., \cite{ding2021copa, gias2019atom, rossi2020hierarchical, tong2021holistic}) highlights the use of queuing theory to estimate scaling metrics (i.e., response time) for different workload levels. In queuing theory, a microservice application is represented as a queuing model, where both service and inter-arrival times follow general statistical distributions \cite{rossi2022dynamic}. However, the accuracy of queuing models can be compromised when there are significant deviations from the exponential distribution in inter-arrival or service times \cite{rossi2020hierarchical, kang2020robust}. Also, queuing models provide approximate estimations, and fine-tuning their parameters requires thorough application profiling, which can be costly and time-consuming \cite{rossi2020self}.
\newline \noindent\textit{iv) Control theory-based scaling policy.} A control theory-based scaling policy \cite{joshi2023arima, baresi2016discrete, baresi2020simulation, baarzi2021showar} fine-tunes system behavior by evaluating the current value of a scaling metric against its reference value within a feedback loop. For example, Baarzi et al. \cite{baarzi2021showar} proposed a proportional–integral–derivative controller \textit{SHOWAR} that leverages the history of scaling decisions and current scaling metrics to formulate the next auto-scaling decisions. Nevertheless, these policies can be time-consuming during their decision-making process, particularly when scaling metrics interact in a complex way because of the inter-dependencies among microservices \cite{rossi2022dynamic}. 
\newline \noindent\textit{v) Artificial Intelligence-based scaling policy.} An Artificial Intelligence (AI) based scaling policy utilizes Machine Learning (ML) and Reinforcement Learning (RL) models to estimate scaling metrics for executing auto-scaling of microservices \cite{toka2021machine}. For regression-based auto-scalers \cite{yang2014cost, islam2012empirical, roy2011efficient}, ML models use historical data of scaling metrics (e.g., CPU and memory usage) for decision-making. However, frequent changes in workload patterns can lead to high costs and time for model retraining \cite{rossi2022dynamic}. For RL-based auto-scalers, RL agents determine scaling actions by engaging in a sequence of interactions with their environment \cite{santos2023gym}. Model-free RL algorithms, such as Q-learning and SARSA \cite{zhang2020sarsa, nouri2019autonomic, horovitz2018efficient}, suffer from slow learning rates. This results in time-consuming auto-scaling of microservices \cite{yu2020microscaler, rossi2020hierarchical}. While model-based RL approaches \cite{rossi2019horizontal, rossi2020self, yang2019miras} can address the slow convergence issue of model-free methods, they face issues related to scalability when applied in large-scale microservice architectures \cite{rossi2022dynamic}.
\vspace{-0.25cm}\subsection{Control Architectures}
Control architecture refers to the structural design of an auto-scaler that is responsible for executing scaling operations based on a scaling policy within microservice applications. In general, control architectures are implemented through MAPE-K control loop components, involving Monitoring, Analysis, Planning, and Execution of auto-scaling operations, facilitated by a Knowledge Base \cite{arcaini2015modeling}. In the existing literature, we have identified two distinct architectural styles for auto-scalers: (i) centralized, and (ii) decentralized.
\newline \indent \textit{Centralized architecture.} The majority of the existing HPAs, such as \cite{gias2019atom, barna2017delivering, khazaei2017elascale}, follow a centralized architectural style for the execution of scaling operations within microservice applications. In this architecture, all data generated by microservices within a microservice application is collected and processed in a central auto-scaler. This central auto-scaler is then responsible for formulating and implementing scaling decisions for all microservices. Although the design of centralized architectures is simple, the centralization of data management leads to several challenges, including the risk of a single point of failure, limited scalability, longer processing times, and a heavier computational load \cite{rossi2020hierarchical}.
\newline \indent \textit{Decentralized architecture.} To address the limitations of centralized architectures, in decentralized architectural style \cite{nitto2020autonomic, nouri2019autonomic}, each microservice within a microservice application is equipped with an independent, dedicated auto-scaler. These individual auto-scalers are responsible for collecting and processing data from their respective microservice and making and executing scaling decisions exclusively for those specific microservices. Moreover, most of the industrial container orchestration platforms (e.g., Kubernetes, Amazon ECS) employ fully decentralized architectures for auto-scaling operations \cite{nguyen2020horizontal}. However, the lack of synchronization among decentralized auto-scalers can lead to frequent scaling operations, particularly in scenarios involving interdependent microservices that contend for resources \cite{rossi2020hierarchical}. This situation degrades the performance of microservice applications. To address this issue, recent works have proposed hierarchical \cite{rossi2020self, rossi2020hierarchical} and master-worker \cite{rossi2020geo, imdoukh2020machine, rossi2019horizontal} decentralized architectural styles that make coordination among independent auto-scalers. However, these approaches introduce communication overhead and bottleneck problems due to increased communication between workers and master auto-scalers during auto-scaling operations \cite{weyns2013patterns}.

\textbf{Distinguishing features of Smart HPA.} The architecture of Smart HPA incorporates both centralized and decentralized components to leverage the strengths of both architectural styles while mitigating their limitations. In resource-rich environments, Smart HPA follows a fully decentralized architectural style, wherein each microservice is equipped with its dedicated auto-scaler, to avoid the limitations of centralized architectures, such as a single point of failure. However, in scenarios where resources are limited, Smart HPA employs a hybrid approach, where it uses a centralized adaptation module for resource management along with decentralized auto-scalers to ensure coordination. It is important to highlight that, unlike existing hierarchical and master-worker architectures, Smart HPA mitigates communication overheads and bottleneck issues by utilizing centralized data processing capabilities only when necessary. Besides, Smart HPA can seamlessly incorporate any scaling policy within its dedicated auto-scalers. While existing auto-scaling policies are bound to make scaling decisions according to the pre-defined resource limits of microservice deployments, Smart HPA enables scaling policies to execute decisions that go beyond the resource limits of microservice deployments. As a prototype implementation, we show the benefits of Smart HPA instantiated with a threshold-based scaling policy due to its straightforward implementation.

\begin{figure*}
  \centering
      \includegraphics[trim=70 30 120 30,clip, width=1.06\linewidth,height=0.565\linewidth]{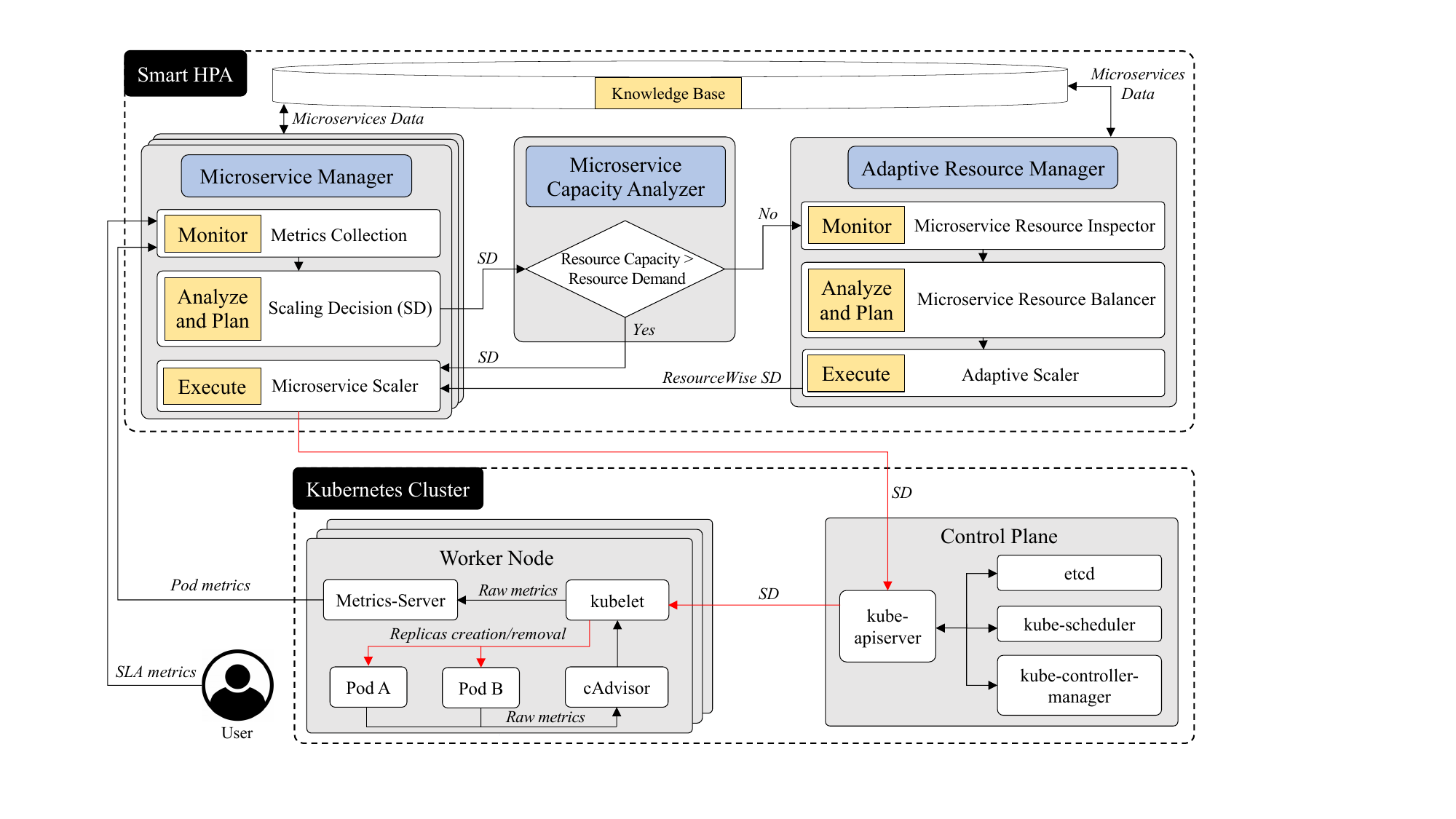}
  \caption{Proposed Hierarchical Architecture of Smart HPA.}
  \label{fig:SmartHPAArchitecture} 
  \vspace{-0.5\baselineskip}  
\end{figure*}
\section{Smart HPA Architecture} \label{section2}
In this section, we describe the hierarchical architecture and resource-efficient heuristics of Smart HPA. Fig.~\ref{fig:SmartHPAArchitecture} presents the hierarchical architectural design of our proposed Smart HPA, which consists of three main components: \textit{Microservice Manager}, \textit{Microservice Capacity Analyzer}, and \textit{Adaptive Resource Manager}. To adapt to frequent resource congestion, Smart HPA incorporates the components of the MAPE-K framework \cite{arcaini2015modeling} into both the decentralized Microservice Managers and the centralized Adaptive Resource Manager for executing the auto-scaling of microservices.

As presented in Fig.~\ref{fig:SmartHPAArchitecture}, Smart HPA employs a dedicated Microservice Manager for each microservice within a microservice application running on a Kubernetes cluster. This dedicated allocation of Microservice Managers offers a high degree of flexibility in auto-scaling operations. For instance, it allows the customization of scaling policies, goals, and metrics tailored to the requirements of each microservice within an application. This provides a clear separation of adaptation goals among individual microservices. Moreover, all Microservice Managers operate in a fully decentralized manner, working in parallel to collect and process data. This decentralized architecture results in enhanced monitoring and improved time efficiency, as opposed to a sequential centralized approach. Initially, all Microservice Managers collect and analyze data from their respective microservices, using a scaling policy to make scaling decisions based on the requirements of each microservice. Subsequently, the Microservice Capacity Analyzer assesses the feasibility of executing scaling decisions by comparing resource demands and resource capacities of all microservices within an application. In resource-constrained situations, where the resource demand exceeds the capacity of a microservice within an application, the Microservice Capacity Analyzer triggers the intervention of the centralized Adaptive Resource Manager. This manager employs resource-efficient heuristics, coordinating Microservice Managers to exchange resources among microservices, and facilitating the formulation and execution of scaling decisions for each microservice. 

It is noteworthy that data generated by the components of Smart HPA for each microservice, such as resource utilization and scaling decisions, is stored within the Knowledge Base of the Smart HPA architecture. The Knowledge Base facilitates further data processing within Smart HPA and enhances situational awareness for key stakeholders, such as developers, practitioners, and users. In the following, we discuss the components of Smart HPA in detail.

\vspace{-0.5\baselineskip}
\subsection{Microservice Manager}
The Microservice Manager is composed of the components of the MAPE-K framework \cite{arcaini2015modeling}. Its primary role involves the collection and analysis of data from a microservice to determine its desired replica count for making a scaling decision accordingly. The Execute component of a Microservice Manager scales a microservice based on instructions from either the Microservice Capacity Analyzer or the Adaptive Resource Manager (Fig.~\ref{fig:SmartHPAArchitecture}). Algorithm \ref{alg:micromanager} outlines the functionality of a Microservice Manager dedicated to a microservice \textit{i}. 

\subsubsection{Monitor}
The Monitor component is responsible for collecting data pertaining to a specific microservice. To make effective scaling decisions, the Monitor component collects essential pod and SLA metrics associated with a microservice. Pod metrics refer to key parameters that reflect a microservice's current status, such as resource utilization (\textit{CMV}) and current replica count (\textit{CR}). Similarly, SLA metrics provide insights into user requirements, including resource threshold value (\textit{TMV}) and minimum/maximum replica counts (\textit{minR, maxR}) of a microservice. Once the data collection phase is complete, the collected metrics are then forwarded to the Analyze component for subsequent data processing.

\subsubsection{Analyze and Plan}

The Analyze and Plan component is responsible for determining the desired replica counts, identifying violations, and making scaling decisions for a microservice.
\afterpage{
\begin{algorithm}[ht]
\caption{\textsc{Microservice Manager}}\label{alg:micromanager}
\small
\begin{tabular}{p{8.25cm}}
  \textit{CMV:} current value of resource metric, 
  \textit{CR:} current replica count,
  \textit{DR:} desired replica count,
  \textit{TMV:} threshold value of resource metric,
  \textit{maxR:} maximum replica count,
  \textit{minR:} minimum replica count,
  \textit{SD:} scaling decision,
  \textit{KB:} knowledge base  \\
  \hline
\end{tabular}
\vspace{5pt}

{\footnotesize$\textbf{Monitor: } CMV_i, TMV_i, CR_i, minR_i, maxR_i $\\
$\textbf{Output: } DR_i, SD_i$}
\begin{algorithmic}[1]
\State $DR_i = \text{ceil}(CR_i \times ({CMV_i} / {TMV_i}))$
\If {$DR_i > CR_i$}
\State $SD_i =  \text{Scale up}$ 
\ElsIf {$DR_i < CR_i \text{ and } DR_i \geq minR_i $}
\State $SD_i =  \text{Scale down}$ 
\Else 
\State $SD_i =  \text{No Scale}$ 
\EndIf
\State $KB \leftarrow DR_i, SD_i, CMV_i,TMV_i, CR_i, minR_i, maxR_i$
\State \Return $DR_i, SD_i, maxR_i$
\end{algorithmic}
\end{algorithm} \vspace{-0.2cm}}
The functionality of the Analyze and Plan  component for a microservice \textit{i} is provided in Algorithm \ref{alg:micromanager} and summarized in the following steps:
\newline \noindent\textit{Step 1: Desired Replicas Determination}. We employ a static threshold-based scaling policy, similar to the one used in Kubernetes HPA \cite{nguyen2020horizontal}, within the Analyze and Plan components of all Microservice Managers to determine desired replica counts for their respective microservices. In Algorithm \ref{alg:micromanager}, the first line illustrates the threshold-based scaling policy, which factors in the current replica count (\textit{$CR_i$}), resource metric utilization (\textit{$CMV_i$}), and resource threshold value (\textit{$TMV_i$}) of a microservice \textit{i} to compute its desired replica count. 
\newline \noindent\textit{Step 2: Violation Detection}. Violations occur when a desired replica count \textit{$DR_i$} for microservice \textit{i} is not equal to its current replica count \textit{$CR_i$}. This indicates the current resource utilization of a microservice is not within its threshold limit. Lines 2 and 4 of Algorithm \ref{alg:micromanager} specify the violation detection conditions and highlight the need for auto-scaling operations.


\noindent\textit{Step 3: Scaling Decision}. Upon detecting a violation, an adaptation process is triggered. This leads to subsequent scale-up or scale-down operations for a microservice. Lines 2-7 of Algorithm \ref{alg:micromanager} outline the scaling decisions \textit{SD} corresponding to different types of detected violations.

Following the scaling decision, the Microservice Manager stores the processed data related to its microservice in the Knowledge Base \textit{KB} (line 9). Moreover, it transmits the scaling decision \textit{SD}, desired replica count \textit{DR}, and maximum replica count \textit{maxR} to the Microservice Capacity Analyzer (line 10).
\vspace{-0.5\baselineskip}
\subsection{Microservice Capacity Analyzer}
When an adaptation is triggered by a Microservice Manager, Smart HPA needs to create or remove replicas for the respective microservice in accordance with its scaling decision (\textit{SD}). As discussed, each microservice within an application is allocated defined resources \cite{zargarazad2023auto}. It is critical to determine whether the resource demand (\textit{DR}) falls within the bounds of the resource capacity (\textit{maxR}) of a microservice. Therefore, we introduce the Microservice Capacity Analyzer, dedicated to assessing the resource demand and resource capacity for each microservice within a microservice application.

As shown in Fig.~\ref{fig:SmartHPAArchitecture}, the Microservice Capacity Analyzer first gathers information on the resource demand and resource capacity for each microservice. Subsequently, in case the resource demand for each microservice aligns with its resource capacity (i.e., \textit{$DR_i \leq maxR_i$}), the Microservice Capacity Analyzer instructs the Execute components of Microservice Managers to proceed with the scaling decisions for their respective microservices. Alternatively, when any microservice, within an application, demands more resources than its resource capacity (i.e., \textit{$DR_i > maxR_i$}), the Microservice Capacity Analyzer activates the centralized component of Smart HPA (i.e., Adaptive Resource Manager) to exchange resources among microservices for executing scaling decisions. This purposeful activation of the centralized Adaptive Resource Manager allows Smart HPA to potentially reduce communication overhead between its centralized and decentralized components.

\begin{algorithm}
\caption{\textsc{Adaptive Resource Manager}}\label{algo:ARM}
\small
\begin{tabular}{p{8.25cm}}
{\footnotesize
  \textit{Overprov:} overprovisioned microservices' resource, 
  \textit{Underprov:} underprovisioned microservices' resource,
  \textit{ResReq:} resource request value for a replica,
  \textit{ResSD:} resource-wise scaling decision,
  \textit{ResDR:} resource-wise desired replica count,
  \textit{TotalR:} total replica count from residual resources} \\
  \hline
\end{tabular}
\vspace{2pt}

{\footnotesize$\textbf{Input: } DR_i, SD_i, maxR_i, ResReq_i$ \hspace*{2.3cm} $\forall i=1,\cdots,M$
$\textbf{Output: } ResSD_i, ResDR_i$ \hspace*{3.5cm} $\forall i=1,\cdots,M $

\Comment{Calculating residual and required resources for all services}}
\begin{algorithmic}[1]
\State \textbf{function} $\textsc{\textbf{Microservice Resource Inspector}}$

\State $Overprov = $[ ], $Underprov = $[ ] 
\For {$i=1 $ to $M$}    \hspace*{0.5cm}        \Comment{M = total number of microservices}
\If {$DR_i > maxR_i$}
\State $RequiredR_i =  DR_i - maxR_i$ 
\State $RequiredRes_i =  RequiredR_i \times ResReq_i$ 
\State $Underprov\text{.append}(RequiredRes_i)$
\Else
\State $ResidualR_i =  maxR_i - DR_i$ 
\State $ResidualRes_i =  ResidualR_i \times ResReq_i$ 
\State $Overprov\text{.append}(ResidualRes_i)$
\EndIf
\EndFor
\State \Return $Underprov, Overprov$

\vspace{2pt}

\noindent\Comment{Resource transfer from overprovision to underprovision services; suggesting feasible desired replicas (\textit{FeasibleR}) and resource capacities (\textit{UmaxR}) for all services}

\vspace{2pt}

\State \textbf{function} $\textsc{\textbf{Microservice Resource Balancer}}$


\State $FeasibleR = $[ ], $UmaxR = $ [ ] 
\State $U = \text{length}(Underprov)$, $O = \text{length}(Overprov)$

\State $TotalOverprov = sum(Overprov)$   \Comment{total residual resource}

\noindent \Comment{Resource reallocation for underprovisioned microservices}

\State $Underprov \gets \text{Dsort}(Underprov)$  \Comment{sort in descending order}

\For {$i=1 $ to $U$}  \hspace*{0.3cm}  \Comment{$U$ = total underprovision microservices}
\State $TotalR_i = {TotalOverprov} / {ResReq_i} $    
\If {$TotalR_i \geq RequiredR_i$}
\State $FeasibleR_i,UmaxR_i = DR_i$ 
\ElsIf {$TotalR_i \in [1, RequiredR_i)$}
\State $FeasibleR_i,UmaxR_i = \text{floor}(TotalR_i) + maxR_i$ 
\Else
\State $FeasibleR_i $, $ UmaxR_i = maxR_i $  
\EndIf
\State $UsedRes_i = (FeasibleR_i - maxR_i) \times ResReq_i$
\State $TotalOverprov = TotalOverprov - UsedRes_i$
\EndFor

\noindent\Comment{Resource reallocation for overprovisioned microservices}

\State $Overprov \gets \text{Asort}(Overprov)$ \Comment{sort in ascending order}

\For {$i=1 $ to $O$}    \Comment{$O$ = total overprovisioned microservices}
\State $TotalR_i = {TotalOverprov} / {ResReq_i}$
\If {$TotalR_i \geq ResidualR_i$}
\State $UmaxR_i = maxR_i$ 
\ElsIf {$TotalR_i \in [1, ResidualR_i)$}
\State $UmaxR_i = \text{floor}(TotalR_i) + DR_i $
\Else
\State $UmaxR_i = DR_i$
\EndIf

\State $FeasibleR_i = DR_i$
\State $UsedRes_i = (maxR_i- UmaxR_i) \times ResReq_i$
\State $TotalOverprov = TotalOverprov - UsedRes_i$
\EndFor
\State \Return $FeasibleR, UmaxR$

\noindent\Comment{Updating desired replica counts, scaling decisions, and resource capacities for all microservices}

\State \textbf{function} $\textsc{\textbf{Adaptive Scaler}}$

\For {$i=1$ to $M$}
\If {$FeasibleR_i == DR_i $}
\State $ResSD_i = SD_i$

\ElsIf {$FeasibleR_i \in (maxR_i, DR_i)$ }


\State $ResSD_i = \text{Scale up}$
\Else                                          
\State $ResSD_i = \text{No Scale}$

\EndIf 
\State $maxR_i = UmaxR_i$, $ResDR_i = FeasibleR_i$
\EndFor
\State $KB \leftarrow Underprov, Overprov, ResSD_i, ResDR_i, maxR_i $ 
\State \Return $ResSD_i, ResDR_i $
\end{algorithmic}
\textbf{end}
\end{algorithm}

\vspace{-0.5\baselineskip}
\subsection{Adaptive Resource Manager} \label{ARM}

The centralized component of our hierarchical Smart HPA, the Adaptive Resource Manager, establishes coordination among decentralized Microservice Managers to execute scaling decisions in resource-constrained environments (i.e., \textit{$DR_i > maxR_i$}). It optimizes resource exchange among microservices to ensure that each microservice's resource demand is adequately met, all while considering the overall resource capacity available for the whole application. We propose resource-efficient heuristics outlined in Algorithm \ref{algo:ARM} for the Adaptive Resource Manager to enable resource-wise scaling of microservices. Fig.~\ref{fig:SmartHPAArchitecture} presents the MAPE-K components of the Adaptive Resource Manager, which are distributed across three key components: Microservice Resource Inspector, Microservice Resource Balancer, and Adaptive Scaler.

\subsubsection{Microservice Resource Inspector}

Lines 1-14 of our heuristics presented in Algorithm \ref{algo:ARM} provide operational details of the Microservice Resource Inspector. To efficiently exchange resources among microservices, it is crucial to determine which microservices have residual resources and which ones are in need of additional resources. Therefore, the Microservice Resource Inspector identifies over- and under- provisioned microservices within a microservice application. It calculates required resources \textit{Underprov} for underprovisioned microservices (lines 4-7) and residual resources \textit{Overprov} for overprovisioned microservices (lines 8-11).

\subsubsection{Microservice Resource Balancer}

The Microservice Resource Balancer transfers resources from overprovisioned to underprovisioned microservices, leading to potential changes in resource capacities (\textit{maxR}) and desired replica counts (\textit{DR}) of microservices. Consequently, the Microservice Resource Balancer suggests feasible desired replica counts (\textit{FeasibleR}) and updated resource capacities (\textit{UmaxR}) for all microservices. The proposed heuristics for the Microservice Resource Balancer are presented in lines 15-46 of Algorithm \ref{algo:ARM}. 

To prioritize addressing the needs of highly underprovisioned microservices (i.e., those experiencing a high load), the Microservice Resource Balancer initiates a process by extracting residual resources from the most heavily overprovisioned microservice, typically the one with the greatest residual resources, and reallocates these resources to the most underprovisioned microservice. Therefore, we sort the underprovisioned microservices' resource (\textit{Underprov}) in descending order (line 19), and the overprovisioned microservices' resource (\textit{Overprov}) in ascending order (line 32). This process iteratively proceeds, starting from the most severely underprovisioned microservice and gradually addressing less underprovisioned ones, until the resource requirements of all underprovisioned microservices are fulfilled (lines 20-31). Consequently, the Microservice Resource Balancer reduces the resource capacities of overprovisioned microservices, indicating retrieval of residual resources from them (lines 33-45). In cases where residual resources from overprovisioned microservices are insufficient to address the demands of underprovisioned microservices, the Microservice Resource Balancer determines the maximum possible desired replica count for underprovisioned microservices (lines 24-25). This is achieved by utilizing the remaining residual resources, thereby maximizing the use of overprovisioned resources to cater to the most pressing needs of underprovisioned microservices. Moreover, in highly resource-constrained situations, where no residual resources are available, no resource exchange takes place (lines 26-27).

\subsubsection{Adaptive Scaler}

Once the Microservice Resource Balancer suggests feasible replica counts (\textit{FeasibleR}) and updated resource capacities (\textit{UmaxR}) for all microservices, the Adaptive Scaler makes scaling decisions and changes resource capacities and desired replica counts accordingly for each microservice within an application (lines 48-57 of Algorithm \ref{algo:ARM}). Here, we denote scaling decisions as resource-wise scaling decisions \textit{ResSD} and desired replica counts as resource-wise desired replica counts \textit{ResDR}. Subsequently, the Adaptive Scaler communicates the \textit{ResSD} and \textit{ResDR} for all microservices to the respective Execute components of Microservice Managers for implementing scaling decisions on corresponding microservices (Fig.~\ref{fig:SmartHPAArchitecture}). Moreover, the Adaptive Scaler stores all data, including \textit{maxR}, \textit{ResSD}, \textit{ResDR}, \textit{Underprov}, and \textit{Overprov}, in the Knowledge Base \textit {KB} of Smart HPA (line 58).

It is important to mention that our proposed heuristics (Algorithm \ref{algo:ARM}) can integrate with any scaling policy (Section \ref{section5}), and metrics, such as CPU usage and response time. This flexibility allows researchers and practitioners to easily choose scaling policies and metrics according to specific requirements.

\section{Experimental Evaluation} \label{section3}
\subsection{Experimental Setup}
\textbf{Experiment Environment.} We use Amazon Web Services (AWS) \cite{amazon} to assess the performance of Smart HPA. We deploy 10 virtual machines (VMs) to host a benchmark microservice application. Each VM is configured as an Amazon Elastic Compute Cloud (Amazon EC2) \textit{t3.medium} instance, equipped with an Intel Xeon Platinum 8000 series processor,  2-core 3.1 GHz CPU, 4GB of RAM, 5Gbps network bandwidth, 5GiB disk size, supporting up to 3 elastic network interfaces and 18 IP addresses. These instances run on the Linux operating system (AL2\_\texttimes{}86\_64) with the EKS-optimized Amazon Linux AMI.  We use Amazon Elastic Kubernetes Service (EKS) \cite{kube} to deploy a benchmark microservice application. The Amazon EKS cluster employs Kubernetes version 1.24, with default AWS VPC network and subnet settings, IPv4 IP cluster family, API server endpoints having both private and public network access, and incorporates add-ons networking features of EKS cluster, such as kube-proxy, CoreDNS, and VPC CNI. Smart HPA is hosted on a local machine, featuring an Intel Corei7 2.60GHz CPU and 16GB RAM. Smart HPA is connected to the application running on AWS EKS through the AWS command-line interface.

\textbf{Benchmark Microservice Application.} Smart HPA can seamlessly integrate with any microservice application running on a Kubernetes cluster, highlighting its flexibility across diverse microservice applications. To  evaluate Smart HPA, we use a real-world microservice benchmark application, Online Boutique \cite{boutique}, as it conforms to the benchmark selection criteria detailed in \cite{aderaldo2017benchmark} and has been widely adopted by the research community, contributing to advancing the state-of-the-art in microservice architectures \cite{santos2023gym, choi2021phpa, karn2022automated}. Online Boutique is a web-based e-commerce application that allows users to browse products, add items to their shopping carts, and make purchases. The application comprises 11 microservices,  implemented in various programming languages.The application also provides a load test script that enables us to assess the scalability of the microservice architectures on which it is deployed. To expedite the deployment process and reduce network bandwidth usage, we have pre-downloaded all the Docker images associated with the Online Boutique onto each VM. 
\begin{figure} [t]
  \hspace*{0.3cm}
  \includegraphics[trim=90 170 190 130,clip, scale = 0.375]{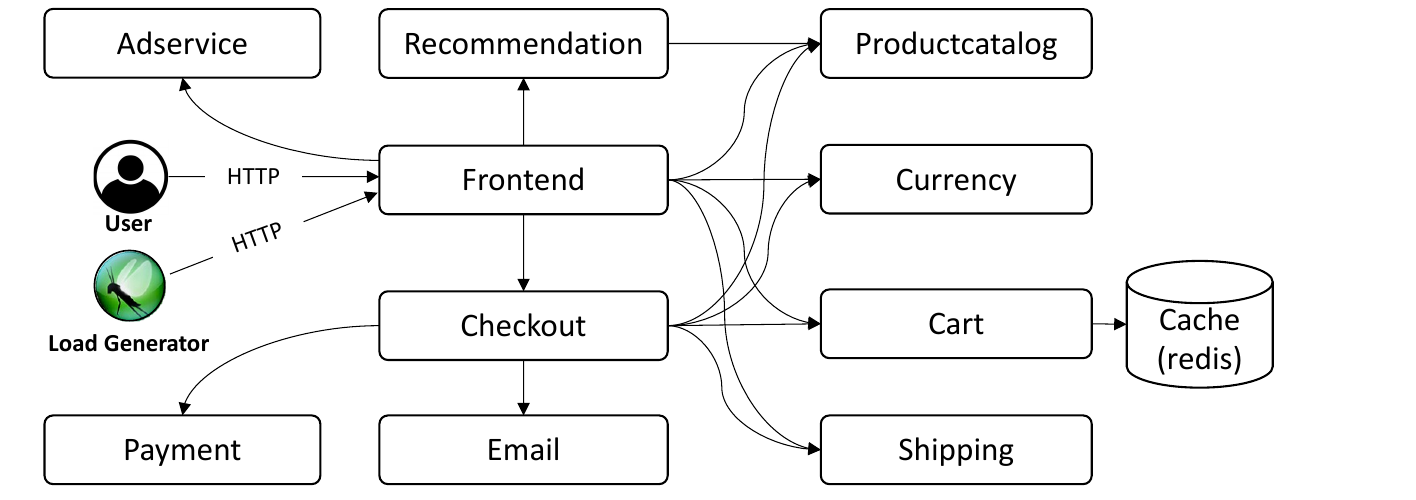}
  \caption{Online Boutique Architecture \cite{boutique}.}
  \label{fig:BenchmarkApp} 
  \vspace{-1.0\baselineskip}
\end{figure}

\begin{figure}[t]
  \centering\vspace{-2mm}
  \begin{minipage}[b]{0.49\linewidth}
    \includegraphics[height=0.8in,width=1.65in]{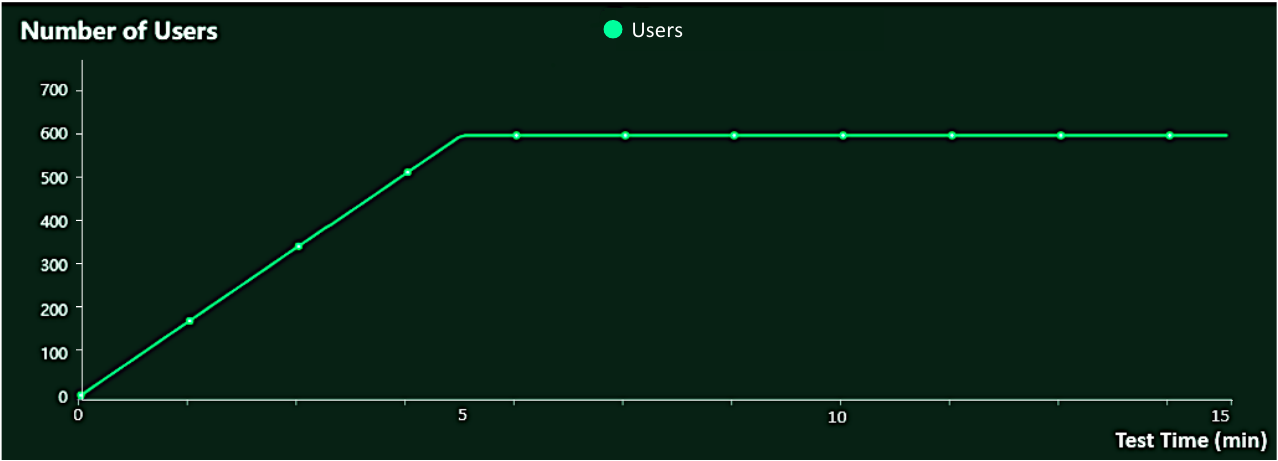}
    \subcaption{Simulated Users.}
    \label{fig:Simulated_Users}
  \end{minipage}
  \hfill
  \begin{minipage}[b]{0.49\linewidth}
    \includegraphics[height=0.8in,width=1.65in]{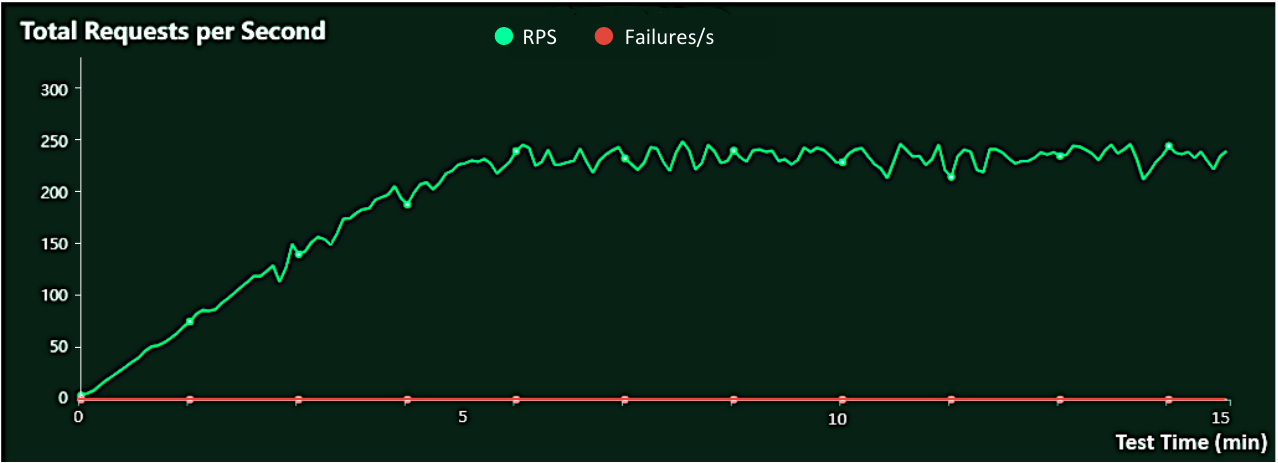}
    \subcaption{Simulated Workload.}
    \label{fig:Simulated_Workload}
  \end{minipage}\vspace{-1.0mm}
  \caption{Load Test for Benchmark Application.}
  \label{fig:LoadTest}
    \vspace{-3mm}
\end{figure}

\textbf{Application Load Testing.} As described earlier, the Online Boutique application provides a load test script to simulate end users for analyzing its scalability. The script simulates user interactions with the benchmark application, such as visiting the homepage, browsing products, adding items to the cart, viewing the cart, checking out, and setting the currency. To execute the load test script, we employ the Locust \cite{locust} load testing tool. We run the Locust on the same local machine where Smart HPA is deployed. As shown in Fig.~\ref{fig:BenchmarkApp}, Locust sends HTTP requests to the benchmark application hosted on AWS EKS. Fig.~\ref{fig:LoadTest} illustrates how Locust simulates users (Fig.~\ref{fig:Simulated_Users}) and the associated workload (Fig.~\ref{fig:Simulated_Workload}) on the benchmark application. The load test is configured to run for a total duration of 15 minutes. In the initial 5 minutes, the test starts with 0 users and gradually increases, simulating the addition of 600 concurrent users with a 2-second spawn rate. This initial phase serves to analyze the behavior of Smart HPA against increasing resource demand. Following this, there are 10 minutes of sustained high load, where all 600 concurrent users actively simulate HTTP requests. This sustained high load creates a resource-constrained scenario for Smart HPA.

\textbf{Evaluation Metrics.} Existing literature has explored a range of resource metrics for executing horizontal pod auto-scaling in microservice architectures. These metrics include, but are not limited to response time, throughput, CPU utilization, traffic load, and memory usage. In this study, we use \textit{CPU utilization} as the scaling metric, aligning with the default metric used by Kubernetes baseline HPA. The common scaling policy and metric selection between Smart HPA and Kubernetes HPA form a solid basis for comparing the two auto-scalers. To assess the hierarchical architecture and resource-efficient heuristics of Smart HPA, we require evaluation metrics that offer insights into resource capacity, demand, shortage, and residual aspects. Therefore, we focus on metrics related to CPU resources outlined in Table \ref{table:EvaluationMetrics}. For example, CPU Underprovision provides insights into required CPU resources, while CPU Overprovision details residual CPU resources. Further details on the evaluation metrics are provided in Table \ref{table:EvaluationMetrics}. It is important to note that Smart HPA offers flexibility to work with other scaling policies and metrics. For instance, if we opt to use response time as a scaling metric and implement a corresponding scaling policy (e.g., \cite{rossi2020hierarchical}) in Microservice Managers, Smart HPA can effectively carry out scaling operations with its resource-efficient heuristics reported in Section \ref{ARM}.

{\footnotesize
\begin{table}[t]
\vspace{-1.0\baselineskip}  
\caption{Evaluation Metrics.} \label{table:EvaluationMetrics}
\begin{tabularx}{88mm}{|>{\hsize=0.3\hsize}X|>{\hsize=0.7\hsize}X|}
\hline

\textbf{Evaluation Metric} & \multicolumn{1}{c|}{\textbf{Description}}\\ \hline

Supply CPU \newline(milliCPU) & CPU resource of current replicas allocated to a microservice. \\ \hline

CPU Overutilization (percent usage) & CPU utilization of a microservice exceeding a predefined threshold value. \\ \hline

Overutilization Time \newline(minutes) & Total duration during which at least one microservice undergoes CPU overutilization. \\ \hline

CPU Overprovision \newline(milliCPU) & The residual CPU resource not utilized by a microservice, (CPU capacity - CPU demand). \\ \hline

Overprovision Time \newline(minutes) & Total duration during which no microservice operates with insufficient CPU resource. \\ \hline

CPU Underprovision \newline(milliCPU) & The CPU resource that a microservice needs but is unavailable, (CPU demand - CPU capacity). \\ \hline

Underprovision Time \newline(minutes) & Total duration during which at least one microservice operates with insufficient CPU resource. \\ \hline
\end{tabularx}
\vspace{-1.5\baselineskip}  
\end{table}
}

\textbf{Experimental Scenarios.} As discussed in Section \ref{section1}, the performance of HPAs is influenced by resource threshold configurations and resource capacities of microservices. 

 \noindent Therefore, to determine the effectiveness of Smart HPA, we have designed experimental scenarios that cover a range of resource capacities and resource threshold configurations.  For resource capacities, we change the maximum number of replicas for each microservice in the benchmark application to simulate various resource-constrained levels for Smart HPA. Specifically, we set resource capacities at 2, 5, and 10 replica counts per microservice. Within each of these settings, we introduce variations in CPU threshold configurations.  In particular, we experiment with CPU threshold configurations set at 20\%, 50\%, and 80\%. This variation in resource capacities and threshold configurations across individual microservices within the benchmark application creates a total of 9 distinct experimental scenarios. We denote each experimental scenario using a combination of the threshold and maximum replica count. For example, we denote the experimental scenario featuring 10 replicas and 50\% CPU Utilization as 10R-50\%.

\begin{figure*}[t]
    \centering

    \begin{subfigure}{.345\linewidth}
        \centering
        \includegraphics[trim={1cm 7cm 0cm 2cm},clip,height=0.5\linewidth,width=1.05\linewidth]{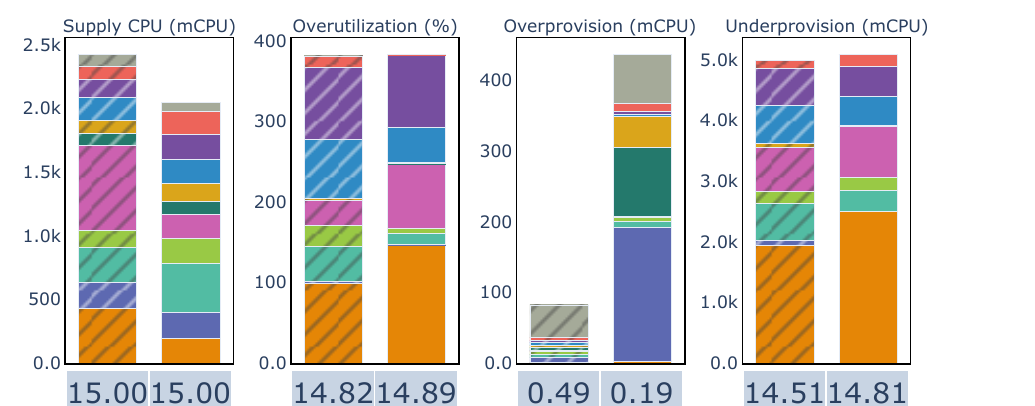}
        \caption{{\scriptsize Max. Replicas: 2 - CPU Threshold: 20\%}}
        \label{fig:2-20}
    \end{subfigure}%
    \begin{subfigure}{.345\linewidth}
        \centering
        \includegraphics[trim={1cm 7cm 0cm 2cm},clip,height=0.5\linewidth,width=1.05\linewidth]{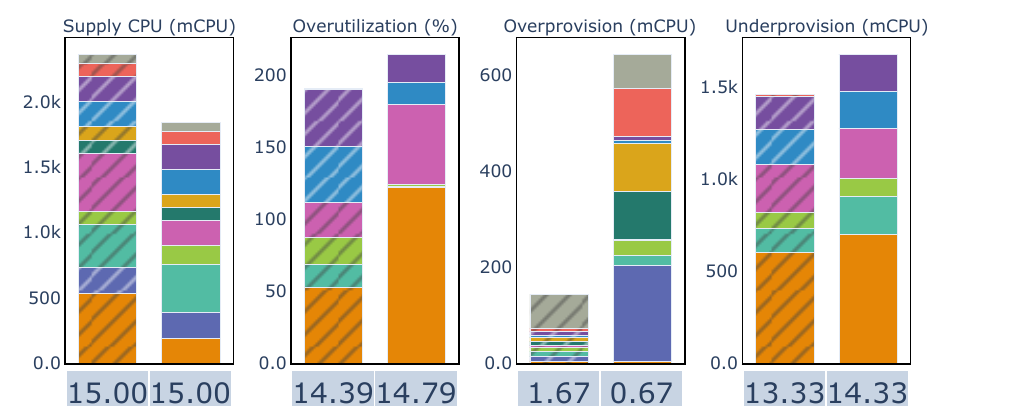}
        \caption{{\scriptsize Max. Replicas: 2 - CPU Threshold: 50\%}}
        \label{fig:2-50}
    \end{subfigure}%
    \begin{subfigure}{.345\linewidth}
        \centering
        \includegraphics[trim={1cm 7cm 0cm 2cm},clip,height=0.5\linewidth,width=1.05\linewidth]{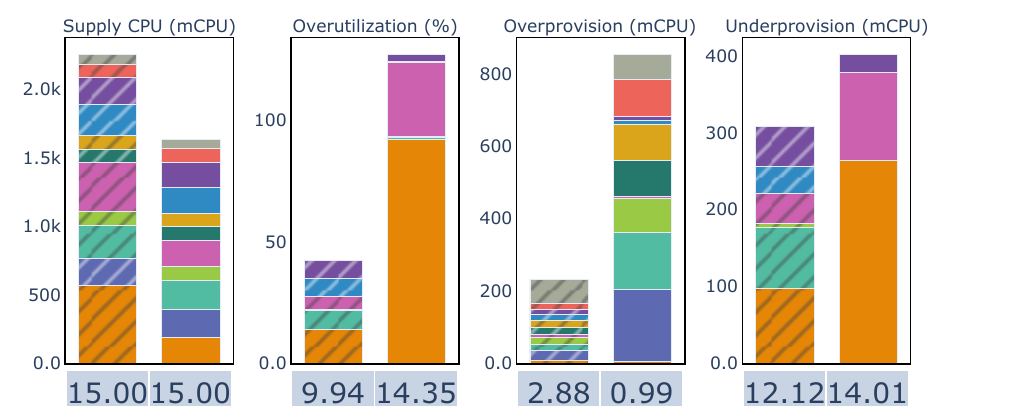}
        \caption{{\scriptsize Max. Replicas: 2 - CPU Threshold: 80\%}}
        \label{fig:2-80}
    \end{subfigure}
    

    \begin{subfigure}{.345\linewidth}
        \centering
        \includegraphics[trim={1cm 7cm 0cm 2cm},clip,height=0.5\linewidth,width=1.05\linewidth]{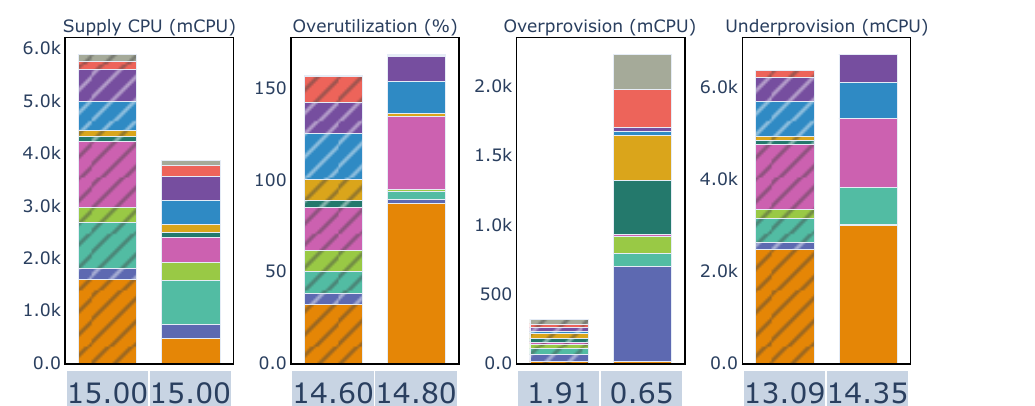}
        \caption{{\scriptsize Max. Replicas: 5 - CPU Threshold: 20\%}}
        \label{fig:5-20}
    \end{subfigure}%
    \begin{subfigure}{.345\linewidth}
        \centering
        \includegraphics[trim={1cm 7cm 0cm 2cm},clip,height=0.5\linewidth,width=1.05\linewidth]{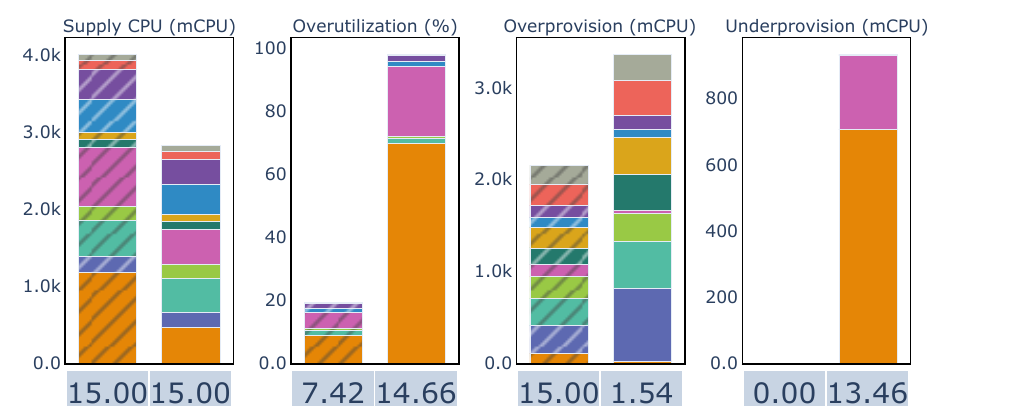}
        \caption{{\scriptsize Max. Replicas: 5 - CPU Threshold: 50\%}}
        \label{fig:5-50}
    \end{subfigure}%
    \begin{subfigure}{.345\linewidth}
        \centering
        \includegraphics[trim={1cm 7cm 0cm 2cm},clip,height=0.5\linewidth,width=1.05\linewidth]{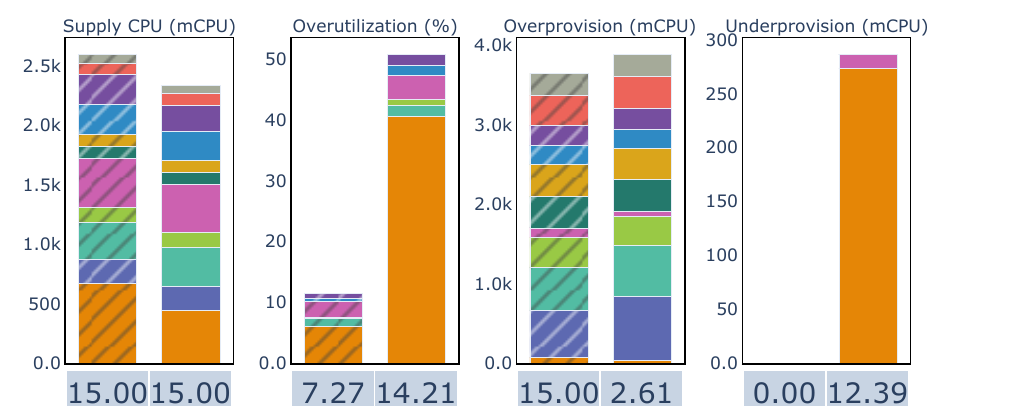}
        \caption{{\scriptsize Max. Replicas: 5 - CPU Threshold: 80\%}}
        \label{fig:5-80}
    \end{subfigure}


    \begin{subfigure}{.345\linewidth}
        \centering
        \includegraphics[trim={1cm 7cm 0cm 2cm},clip,height=0.5\linewidth,width=1.05\linewidth]{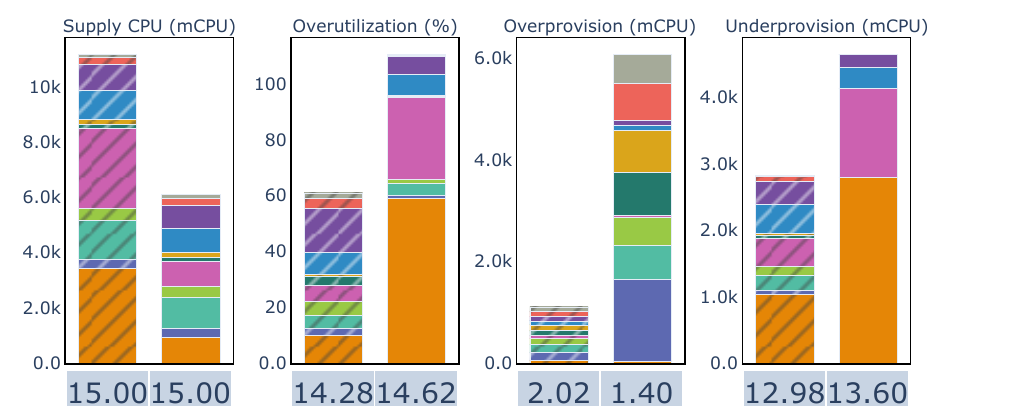}
        \caption{{\scriptsize Max. Replicas: 10 - CPU Threshold: 20\%}}
        \label{fig:10-20}
    \end{subfigure}%
    \begin{subfigure}{.345\linewidth}
        \centering
        \includegraphics[trim={1cm 7cm 0cm 2cm},clip,height=0.5\linewidth,width=1.05\linewidth]{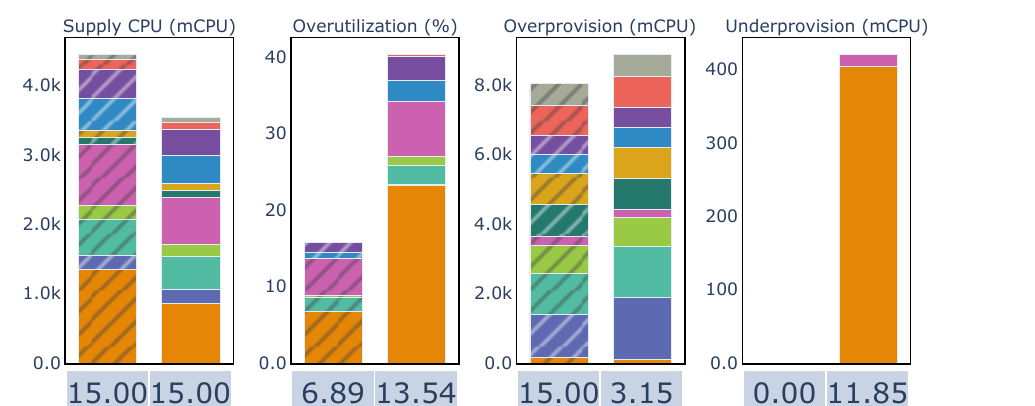}
        \caption{{\scriptsize Max. Replicas: 10 - CPU Threshold: 50\%}}
        \label{fig:10-50}
    \end{subfigure}%
    \begin{subfigure}{.345\linewidth}
        \centering
        \includegraphics[trim={1cm 7cm 0cm 2cm},clip,height=0.5\linewidth,width=1.05\linewidth]{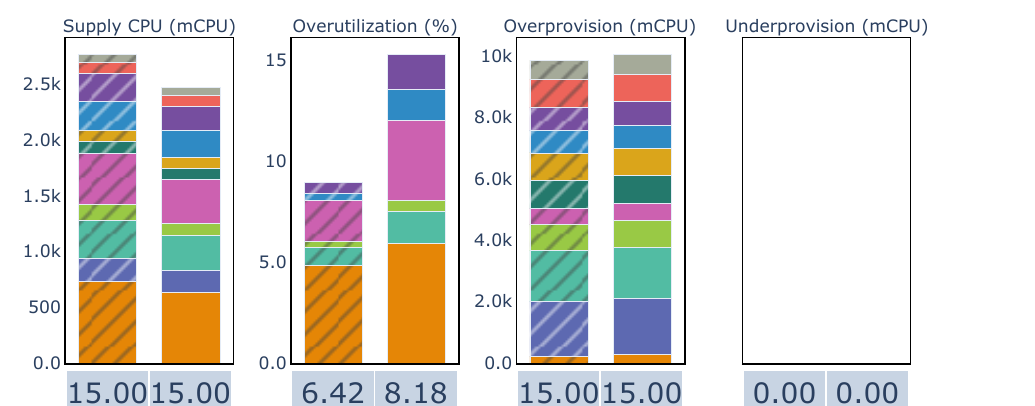}
        \caption{{\scriptsize Max. Replicas: 10 - CPU Threshold: 80\%}}
        \label{fig:10-80}
    \end{subfigure}

\centering
    \includegraphics[trim={0.03cm 0.03cm 0.03cm 0.03cm},clip, width=1\linewidth]{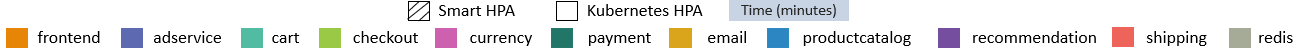}
    \caption{Results for all Experimental Scenarios: Lower values, better performance (except Supply CPU and Overprovision Time).}
    \label{fig:AllResults}
    \vspace{-1.2\baselineskip}  
\end{figure*}

\begin{figure*}[t]
\begin{subfigure}{.50\linewidth}
    \includegraphics[trim={0.3cm 0.8cm 1.5cm 2.6cm},clip,width=1.02\linewidth,height=0.55\linewidth]{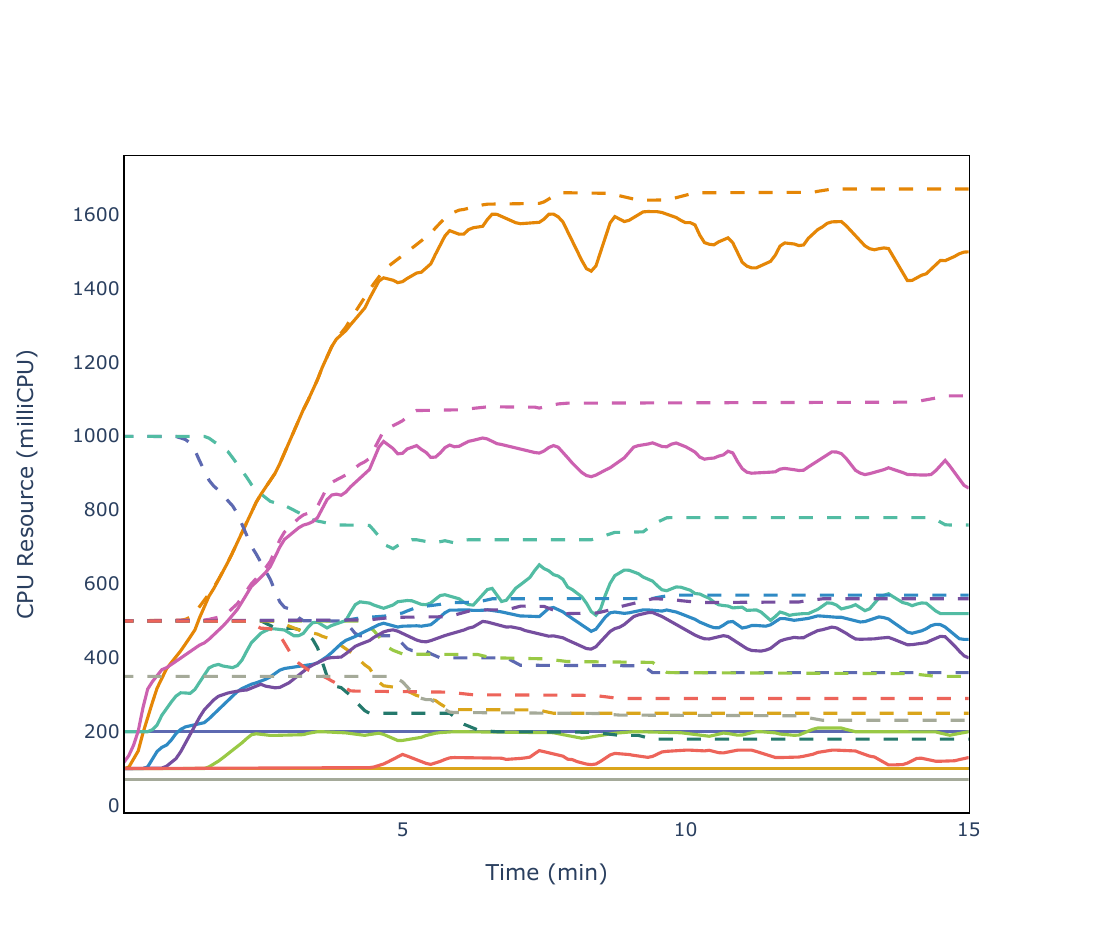}\vspace{-0.2cm}
    \caption{{\footnotesize Smart HPA - CPU Demand vs CPU Capacity}} \label{SmartCap}
\end{subfigure}%
\begin{subfigure}{.50\linewidth}
    \includegraphics[trim={0.3cm 0.8cm 1.5cm 2.6cm},clip,width=1.02\linewidth,height=0.55\linewidth]{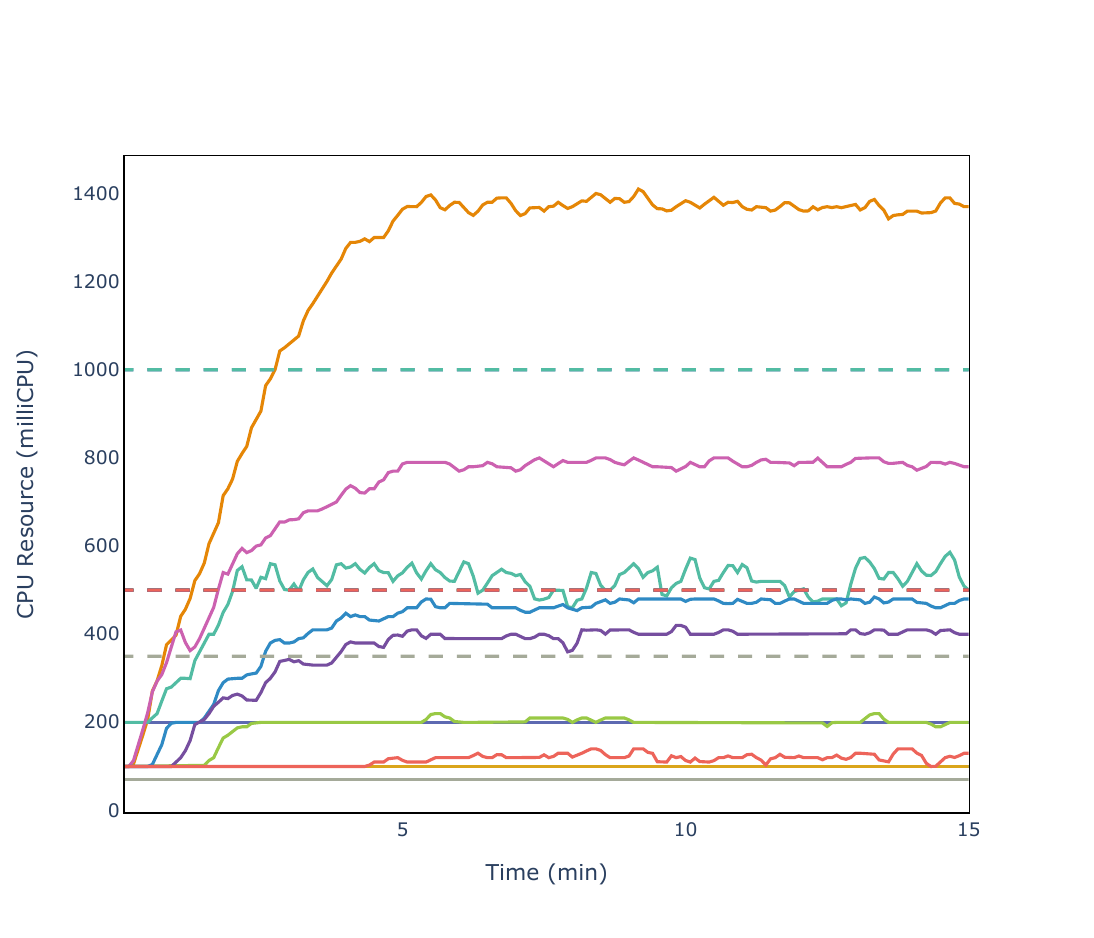}\vspace{-0.2cm}
    \caption{{\footnotesize Kubernetes HPA - CPU Demand vs CPU Capacity}} \label{KubeCap}
\end{subfigure} \vspace{-0.07cm}
\hspace{0.09cm} \begin{subfigure}{.50\linewidth}
    \includegraphics[trim={0.5cm 0.8cm 1.5cm 2.6cm},clip,width=1.00\linewidth,height=0.55\linewidth]{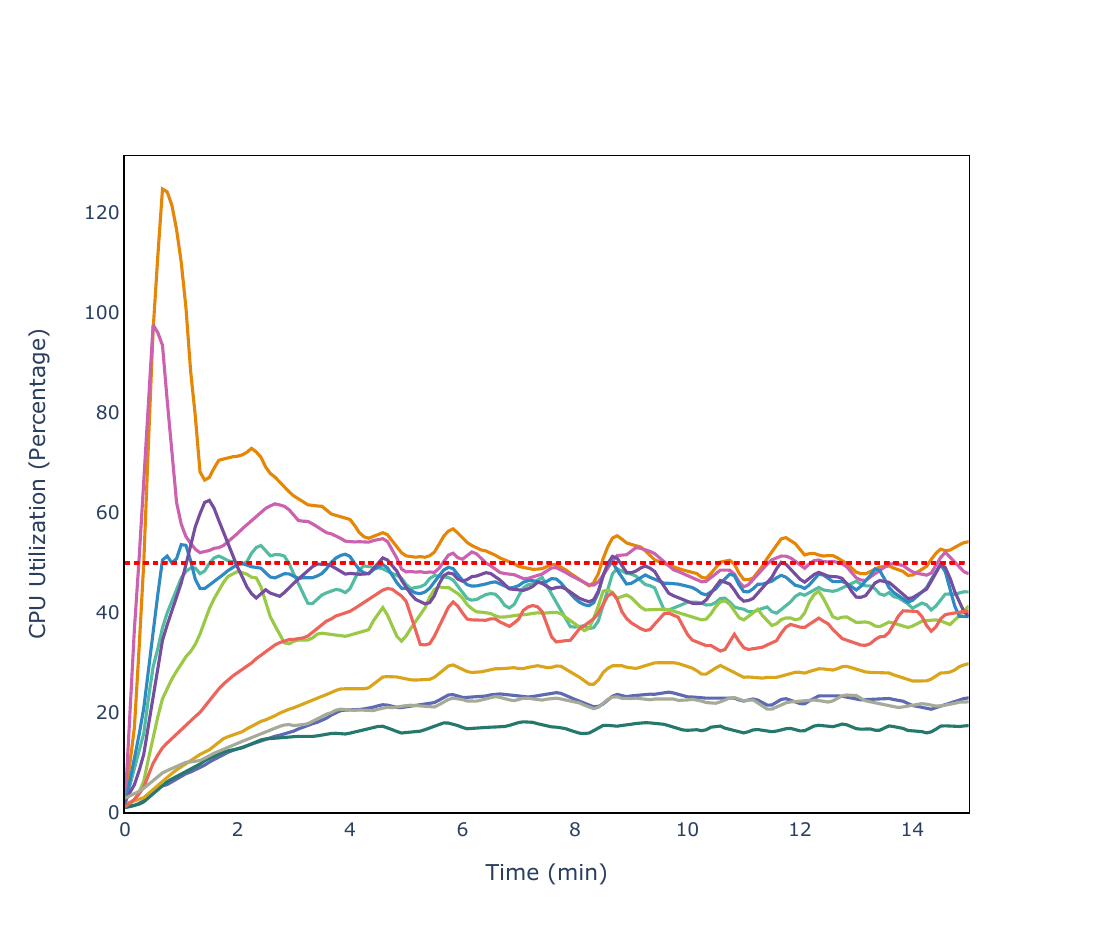}\vspace{-0.2cm}
    \caption{{\footnotesize Smart HPA - Percent CPU Utilization}} \label{SmartUti}
\end{subfigure}%
\begin{subfigure}{.50\linewidth}
    \includegraphics[trim={0.5cm 0.8cm 1.5cm 2.6cm},clip,width=1.00\linewidth,height=0.55\linewidth]{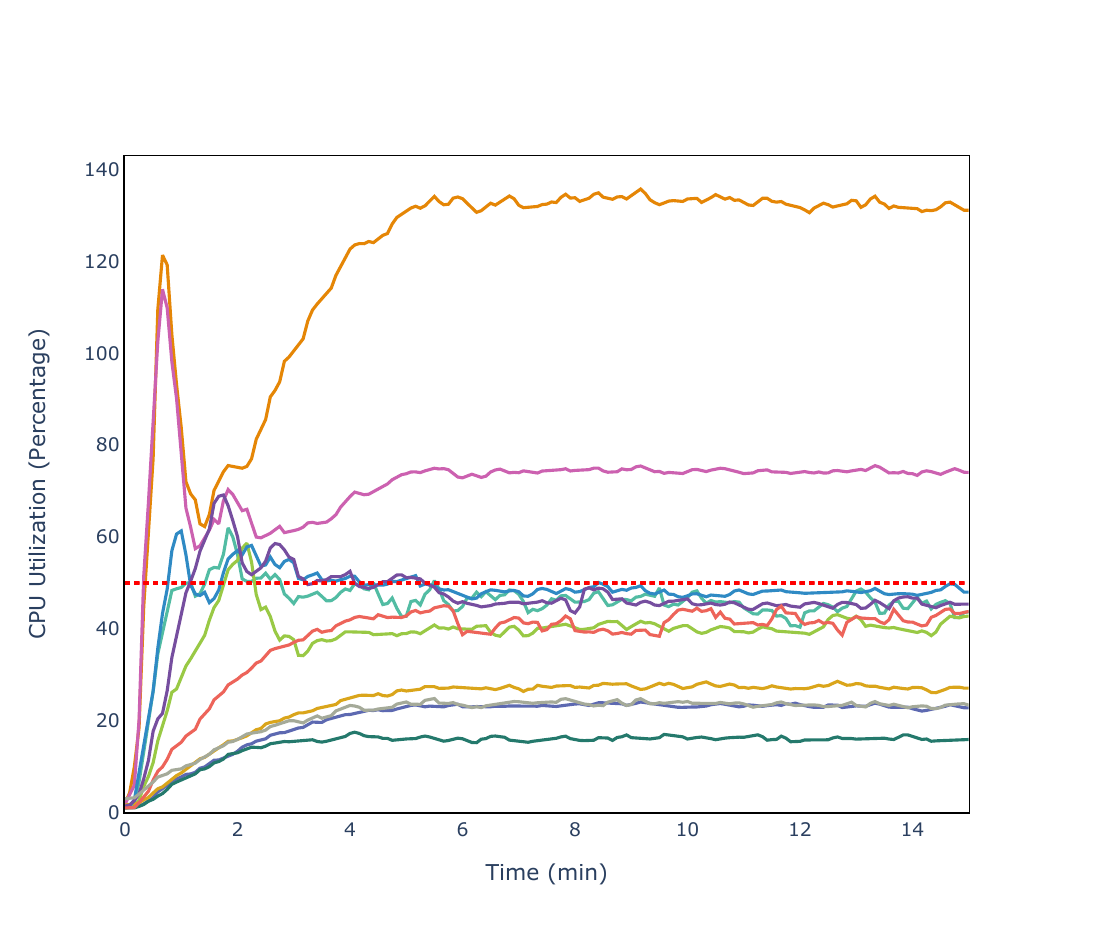}\vspace{-0.2cm}
    \caption{{\footnotesize Kubernetes HPA - Percent CPU Utilization}} \label{KubeUti}
\end{subfigure}
\vspace{-0.2cm}\includegraphics[trim={0.3cm 0.5cm 0.2cm 0.7cm},clip,height=0.05\linewidth,width=1\linewidth]{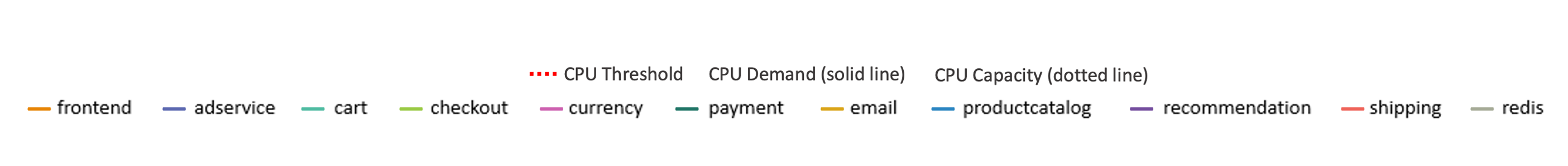}
\caption{Smart HPA vs. Kubernetes HPA: CPU Utilization, CPU Demand, and CPU Capacity for the Scenario 5R-50\%.} \label{SmartvsKub}
\vspace{-1.5\baselineskip}  
\end{figure*}
\vspace{-0.5\baselineskip}
\subsection{Results and Discussion}
We compare the performance of Smart HPA against the Kubernetes baseline HPA. For each experimental scenario, we execute our load test on the benchmark application with Smart HPA and Kubernetes HPA separately. To ensure the reliability of our findings, we conduct each load test 10 times for every scenario and subsequently calculate average results. The benchmark application is configured with default settings, where each replica of all microservices has a CPU request of 100m and a CPU limit of 200m, except for \textit{adservice} and \textit{cart} service, which have a CPU request of 200m and a CPU limit of 300m, and \textit{redis} service with a CPU request of 70m and a CPU limit of 125m. In Fig.~\ref{fig:AllResults}, we present the performance evaluation results for Smart HPA and Kubernetes HPA across all experimental scenarios. The figure depicts results at both the application level and microservice level, with different colors representing the 11 microservices within the benchmark application. We observe that Smart HPA consistently outperforms Kubernetes HPA across all resource levels, ranging from 2 to 10 replicas, and threshold settings spanning from 20\% to 80\% CPU utilization.

\textbf{Smart HPA vs. Kubernetes HPA: Analyzing Optimal and Challenging Scenarios.} We analyze the most and least favorable scenarios for Smart HPA compared to Kubernetes HPA in terms of evaluation metrics.
\newline\textit{Overutilization and Underprovision CPU and Time.} In the scenario 5R-50\% (Fig.~\ref{fig:5-50}), Smart HPA shows no CPU Underprovision, while Kubernetes HPA records 934.04m CPU Underprovision for 13.46 minutes. This marks the highest improvement in both CPU Underprovision and Time compared to all nine scenarios, attributed to the efficient CPU resource balancing of Smart HPA. Consequently, Smart HPA results in 19.30\% CPU Overutilization for 7.42 minutes, compared to Kubernetes HPA's 98.06\% CPU Overutilization for 14.66 minutes, indicating a significant 5.08x reduction in CPU Overutilization and a 1.98x decrease in Overutilization Time compared to Kubernetes HPA. Conversely, in the scenario, 2R-20\% (Fig.~\ref{fig:2-20}), where a lower CPU threshold and fewer replicas contribute to increased resource scarcity, Smart HPA demonstrates only a marginal 1.004x reduction in CPU Overutilization and Time, and a 1.01x reduction in CPU Underprovision and Time compared to Kubernetes HPA.
\newline \noindent\textit{Overprovision CPU and Time.} Low CPU thresholds enable auto-scalers to rapidly scale microservices to their maximum replica counts during high workloads. This gives Smart HPA a performance advantage in reducing CPU Overprovision by exchanging resources among microservices to align the resource capacities of microservices with their resource demands. For instance, in scenarios 2R-20\%, 5R-20\%, and 10R-20\%, Smart HPA demonstrates 5.29x, 7.07x, and 5.46x reduction in CPU Overprovision, respectively, compared to Kubernetes HPA. Regarding Overprovision Time, Smart HPA significantly outperforms Kubernetes HPA in scenarios where the benchmark application experiences zero CPU Underprovision. For example, in scenarios 5R-50\%, 5R-80\%, and 10R-50\%, Smart HPA exhibits 9.74x, 5.75x, and 4.76x increase in Overprovision Time, respectively, compared to Kubernetes HPA. On the other hand, in scenario 10R-80\% (Fig.~\ref{fig:10-80}), where no microservice encounters CPU underprovisioning under both Smart HPA and Kubernetes HPA, Smart HPA exhibits only a minimal 1.01x reduction in CPU Overprovision and no improvement in Overprovision Time compared to Kubernetes HPA.
\newline \noindent \textit{Supply CPU.} Smart HPA allocates the residual capacity of overprovisioned microservices to underprovisioned ones, resulting in a higher supply of CPU resources compared to Kubernetes HPA to the benchmark application. With a low CPU threshold triggering auto-scalers to generate more microservice replicas (i.e., Supply CPU) in response to high workloads, in scenario 10R-20\% (Fig.~\ref{fig:10-20}), Smart HPA supplies 1.83x more CPU resources (11188.76m) to the benchmark application compared to the 6110.41m Supply CPU provided by Kubernetes HPA. Conversely, in scenario 10R-80\% (Fig.~\ref{fig:10-80}), where no CPU underprovisioning occurs, Smart HPA holds a slight performance edge over Kubernetes HPA, supplying 1.11x more CPU resources (2771.42m) compared to 2478.62m CPU resource provided by Kubernetes HPA to the benchmark application during the 15-minute load test.
\newline In summary, Smart HPA's dynamic architecture allows it to outperform  Kubernetes HPA under high workload scenarios.
\newline \textbf{Smart HPA vs. Kubernetes HPA: Adaptive Behaviour Comparison.} To understand the comparative behavior of Smart HPA and Kubernetes HPA during the load test, we present an entire scenario 5R-50\% in Fig.~\ref{SmartvsKub}. The figure depicts the evolution in CPU utilization, CPU demand, and CPU capacity for each microservice throughout the load test. 
At the start of the test, we observe that microservices start using CPU resources as the workload increases, as depicted in Fig.~\ref{SmartUti} and \ref{KubeUti}. The rise in CPU utilization of microservices leads to an increase in their CPU demand, a pattern observed for both Smart HPA (Fig.~\ref{SmartCap}) and Kubernetes HPA (Fig.~\ref{KubeCap}). 

In the case of Smart HPA, as shown in Fig.~\ref{SmartCap}, around 1.50 minutes into the test, the CPU demand for the \textit{frontend} exceeds its allocated CPU capacity of 500m. At this point, the Microservice Capacity Analyzer triggers the Adaptive Resource Manager. The manager identifies the \textit{adservice} as the most overprovisioned, with 1000m CPU resources, and transfers some of its CPU resources to address the underprovisioned state of the \textit{frontend}, ensuring that its capacity matches its demand. As a result, the capacity of the \textit{frontend} increases to meet its increasing demand, while the capacity of the \textit{adservice} decreases but remains above its demand (Fig.~\ref{SmartCap}). Similarly, when the \textit{currency} experiences a resource shortage around 2 minutes into the test, the Adaptive Resource Manager allocates the necessary CPU resources from the most overprovisioned \textit{cart} microservice. When overprovisioned microservices like \textit{adservice} and \textit{cart} can no longer provide additional resources due to their capacity approaching their demand, the Adaptive Resource Manager reallocates CPU resources from other overprovisioned microservices, such as \textit{email} and \textit{shipping}, to address the underprovisioned state of the \textit{frontend} and \textit{currency}. In this way, Smart HPA effectively prevents microservices from becoming underprovisioned. Consequently, as illustrated in Fig. \ref{SmartUti}, CPU utilization of both the \textit{frontend} and \textit{currency} experiences a decline and maintains a closer proximity to the 50\% CPU threshold value.

In contrast, Kubernetes HPA does not facilitate resource sharing among microservices (Fig. \ref{KubeCap}). This results in constant capacity levels (i.e., 500m and 1000m) for microservices throughout the load test. Consequently, this leads to a shortfall of CPU resources for the \textit{frontend} and \textit{currency}. Hence, during high workload, the CPU utilization of both the \textit{frontend} and \textit{currency} remains around 130\% and 70\%, respectively, surpassing the 50\% CPU threshold value (Fig.~\ref{KubeUti}). Therefore, when we compare Smart HPA to Kubernetes HPA (Fig.~\ref{fig:5-50}), Smart HPA significantly reduces CPU Overutilization and Overutilization Time by 5.08x and 1.98x, respectively. It also decreases CPU Overprovision by 1.56x and increases Overprovision Time by 9.74x compared to Kubernetes HPA. Notably, Smart HPA experiences no CPU Underprovision, while Kubernetes HPA encounters 934.04m CPU Underprovision for 13.46 minutes.
Thus, Smart HPA's resource exchange prevents microservices from underprovisioning, resulting in superior performance across all metrics compared to Kubernetes HPA.

\textbf{Implications of Variable Resource Settings.}
\newline\noindent\textit{Implications of changing CPU threshold settings.} We notice a consistent pattern in our evaluation metrics for both Smart HPA and Kubernetes HPA with changing CPU threshold configurations (i,e., 20\%, 50\%, and 80\%), regardless of the number of replicas allocated to microservices. As illustrated in Fig.~\ref{fig:AllResults}, we observe a decrease in Supply CPU as the threshold value rises. For instance, in scenarios 5R-20\%, 5R-50\%, and 5R-80\%, Smart HPA decreases the Supply CPU from 5887.53m to 4013.82m and further to 2601.18m, respectively. This suggests that, as the CPU threshold increases, microservices operate effectively with fewer replicas. Furthermore, with increasing CPU threshold values, Fig.~\ref{fig:AllResults} demonstrates a decreasing trend in CPU Overutilization, CPU Underprovision, and their respective time metrics, accompanied by an increase in CPU Overprovision and its associated time metric. This is due to the fact that each replica of microservice deployments provides more CPU capacity as the CPU threshold increases, which results in reduced CPU Overutilization and CPU Underprovision, and an increase in CPU Overprovision.
\newline\textit{Implications of changing resource capacities.} With an increasing number of replicas (i.e., 2, 5, and 10), Fig.~\ref{fig:AllResults} exhibits an increase in Supply CPU, CPU Overprovision, and Overprovision Time for both Smart HPA and Kubernetes HPA. This increase results from the additional CPU capacity stemming from the increased number of replicas. For example, in scenarios 2R-20\%, 5R-20\%, and 10R-20\%, Smart HPA exhibits a rising trend in CPU Overprovision, with corresponding values of 82.67m, 315.40m, and 1110.91m (Fig.~\ref{fig:AllResults}). As a result of this increased CPU capacity, we observe a decrease in CPU Overutilization, CPU Underprovision, and their associated time metrics. 
\newline\textit{Implications of changing both CPU threshold and resource capacities.} When we investigate the combined impact of altering both threshold configurations and the number of replicas in microservice deployments, we notice that the Supply CPU tends to be higher when the threshold is low and the number of replicas is high. For instance, as illustrated in Fig.~\ref{fig:AllResults}, scenario 10R-20\% records the highest Supply CPU of 11188.76m for Smart HPA and 6110.41m for Kubernetes HPA among all scenarios. This occurs because a low CPU threshold triggers auto-scalers to generate more microservice replicas in reaction to a high workload. Moreover, as both the CPU threshold and the number of replicas increase, CPU capacity also rises, leading to higher CPU Overprovision and Time metrics. Simultaneously, this leads to reduced CPU Overutilization, Underprovision, and their respective Time metrics. Therefore, scenario 10R-80\% has the highest CPU Overprovision (9864.39m) and Overprovision Time (15 minutes) and the lowest CPU Overutilization (8.93\%) and Overutilization Time (6.42 minutes), with zero CPU Underprovision and Time for Smart HPA. 
\newline In summary, increasing CPU threshold or capacity for microservices decreases CPU overutilization and underprovisioning, while increasing CPU overprovisioning, and vice versa.
\vspace{-0.5\baselineskip}
\section{Threats to Validity}
We identify some external threats that could impact the generalizability of our findings. One such threat stems from the use of only one microservice benchmark application for the evaluation, potentially limiting the application of our findings to other microservice-based systems. However, given the widespread use, heterogeneity, and size of the selected benchmark application, we believe that our results can be applicable to other microservice applications or real-world settings. Additionally, the initial resource configurations of microservices, such as resource request and limit values, play a crucial role in the claimed improvement margin in our study. While we observed improvements in resource utilization using the benchmark application with default resource configurations, variations in initial resource configurations could result in smaller or larger improvements. Another external threat emerges from the comparison of our study with the Kubernetes baseline HPA to substantiate our findings, instead of multiple available alternative HPAs. However, the widespread adoption of Kubernetes in both industrial and academic settings mitigates this concern, providing a solid foundation for validating our study. Lastly, a potential threat lies in the variation of Smart HPA behavior under different workload profiles. The selected workload profile follows a pattern of increasing and sustained high workloads, crucial for creating resource-constrained scenarios for HPAs. Therefore, Smart HPA has the potential for consistent performance across a range of workload profiles, including those capable of inducing resource-constrained scenarios.

\vspace{-0.5\baselineskip}
\section{Conclusion and Future Work} \label{section6}
We introduce Smart HPA, featuring a hierarchical architecture that integrates both centralized and decentralized architectural styles to perform horizontal auto-scaling in microservices. Within this hierarchical architecture, decentralized managers monitor microservice resource metrics, like CPU usage, while the centralized manager is selectively activated in resource-constrained environments to efficiently transfer resources between microservices, minimizing communication overhead. In our experiments, Smart HPA outperforms Kubernetes HPA with default configurations for a benchmark application. It reduces resource overutilization by 5x, overutilization time by 2x, and overprovisioning by 7x. Additionally, it eliminates underprovisioning, improves resource allocation by 1.8x, and extends overprovisioning time by 10x.

We have identified several areas for improvement that we plan to address in our future work. One such avenue involves employing AI-based predictive methods, such as time series analysis of workload and resource utilization. This enables Smart HPA to operate with both proactive and reactive auto-scaling mechanisms. Moreover, evaluating Smart HPA with alternative scaling policies, such as queuing theory-based approaches, and assessing metrics like response time and communication overhead will strengthen and validate its flexibility. Furthermore, during our experiments, we observed the startup time of microservice containers significantly impacts the efficiency of auto-scaling operations. As such, a promising avenue for future research involves reducing startup time to expedite auto-scaling in microservice architectures.

\bibliographystyle{ieeetr}
\bibliography{ref}
\end{document}